Title: Yukawa's short-range nuclear force vs. Debye's electrostatic screening
Author: Mladen Georgiev (Institute of Solid State Physics, Bulgarian Academy of
   Sciences, 1784 Sofia, Bulgaria)
Comments: 8 pdf pages with 1 table and 3 figures
Subj-class: physics

The eigenvalue problem of a short-range potential is revisited in view of the increased interest in a simple model imitating the nuclear forces. This is in order to perform calculations of vibronic energies in fermion-boson coupled systems.

1. Foreword

We have recently suggested that coupled fermion-boson systems may be akin to coupled electron-phonon systems in many respects including the behavior of their vibronic potentials as well as the possibility to give rise to symmetry breaking Jahn-Teller effects. In the pursuit of simple-model short-range potentials to imitate the nuclear forces, historically Yukawa's potential appeared the most promising one. A few attempts have been made in the past to deduce eigenstates and eigenvalues for the Yukawa potential, though most of them bear the character of approximations devoid of clear-cut conclusions. In what follows, we also present a state-of-the-art outlook of the formally related problem of Debye screening in electrostatics.

2. Intranuclear potential

Yukawa [1] introduced a phenomenological short-range intranuclear potential by analogy with Debye's screening electrostatic potential. In spherical symmetry:

$$V(r) = - (1/r) \exp(-\kappa_Y r) \qquad (1)$$

Here $\kappa_Y = 5 \times 10^{12}$ cm$^{-1}$ is a constant which has been deduced from the experimental data. It is not clear whether $\kappa$ is an actual constant or is dependent on other parameters. The same potential is characteristic of Debye's screening only by substituting $\kappa_S$ for $\kappa_Y$. To the contrary the respective electron screening constant $\kappa_S = \sqrt{(8\pi n e^2/\varepsilon k_B T)}$ depends on the temperature T and the carrier concentration n. The status of $\kappa$ is thus a point of distinction between Yukawa's and Debye's arguments virtually leading to the same math result. In the following $\kappa \equiv \kappa_Y$ (Yukawa force) and $\equiv \kappa_S$ (Debye screening) under the same letter $\kappa$ will be implied if not mentioned otherwise.

3. Poisson-Boltzmann equation vs. Coulomb screening

The Yukawa potential [1] is akin to the screened Coulomb potential in electrostatics [2]. Both should be dealt with by similar methods. We remind that the latter appears as an approximate easy-to-tackle version at $e\phi \ll k_B T$ (Debye-Hückel's (D-H) approximation) of the electrostatic potential $\phi$ in a Maxwell-Boltzmann (M-B) system [3]. In considering the quantum-mechanic behavior of M-B systems, the complete electrostatic functions at $e\phi \sim k_B T$ should be added and only then will the D-H approximation be employed at the final stage. The electrostatic potential ϕ of a M-B structure is obtained by combining Poisson's equation with Boltzmann's statistics resulting in the Poisson-Boltzmann (P-B) equation:

$$\Delta(e\phi/k_BT) = (4\pi n_0 e^2/\varepsilon k_BT)[\exp(+e\phi/k_BT) - \exp(-e\phi/k_BT)] = (8\pi n_0 e^2/\varepsilon k_BT)\sinh(e\phi/k_BT) \quad (2)$$

or, in a reduced form [$\Phi = e\phi/k_BT$, $\kappa = \sqrt{(8\pi n_0 e^2/\varepsilon k_BT)}$]:

$$\Delta(\Phi) = \kappa^2 \sinh\Phi \quad (3)$$

The non-approximate M-B functions have frequently been used in physics, including nuclear matter and solid state physics [4]. The electrostatic potential is obtained as solution to the P-B equation. There are an unlimited number of solutions depending on the symmetry configuration: 1-D plane-wave, 2-D planar, 3-D spatial, spherical, cylindrical, etc. In the simplest 1-D case there is a minimal-frequency solution to (2):

$$\Phi(\kappa x) = \ln[\cotan(\tfrac{1}{2}\kappa x)]^2 \quad (4)$$

The D-H approximation obtains at $\Phi \ll 1$ in (3) which gives:

$$\Delta(\Phi) = \kappa^2 \Phi \quad (5)$$

In 1-D the solution to (5) is the well-known Debye exponential:

$$\Phi(\kappa x) \sim \exp(-\kappa x) \quad (6)$$

Accordingly, the spherical-symmetry D-H potential drops away from the center at $r = 0$ as

$$\Phi(r) = (1/r)\exp(-\kappa r) \quad (7)$$

This potential is the analogue of the generator of Yukawa short-range force. The aperiodic and periodic 1-D potentials are depicted in Figure 1 and 2, respectively.

### 3. Yukawa equation

Deductions will be made of approximate Yukawa eigenfunctions aimed ultimately at solving the eigenvalue equation. In electrostatic problems, we set $C = q^2$ (q is the electric charge on the particle).

1. We rewrite the respective Schrödinger's equation

$$(-\hbar^2/2M)\Delta\psi - (C/r)\exp(-\kappa r)\psi = E\psi \quad (8)$$

by using the substitution $\psi = [u(r)/r]\, Y_l^m(\theta,\varphi)$ leading to the radial equation:

$$(-\hbar^2/2M)[d^2/dr^2 - l(l+1)/r^2]u - (C/r)\exp(-\kappa r)u = Eu \quad (9)$$

Setting $\exp(-\kappa r) \sim 1 - \kappa r$ at $\kappa r \ll 1$ we simplify the exponential to get accordingly

$$(-h^2/2M)[d^2/dr^2 - l(l+1)/r^2] u - (C/r)u = [E - C\kappa]u \qquad (10)$$

For C < 0, the eigenstates of (10) are the Coulomb functions [5], C > 0 leads to hydrogen eigenstates.

2. We introduce the substitution $\psi(r) = \Psi(r)\exp(-\kappa r)$ into Schrödinger's equation (8) for $\psi$ at $l = 0$ to get an equation for $\Psi$ leading to

$$-(h^2/2M)[\Psi'' - 2\kappa\Psi'] + (C/r)\exp(-\kappa r)\Psi = [E - (h^2\kappa^2/2M)]\Psi \qquad (11)$$

At $\kappa r \ll 1$, $\exp(-\kappa r) \sim 1 - \kappa r$ and then we ultimately get in lieu of the above:

$$-(h^2/2M)[\Psi'' - 2\kappa\Psi'] + (C/r)\Psi = [E - (h^2\kappa^2/2M) - C\kappa]\Psi \qquad (12)$$

3. We rewrite equation (9) to get

$$(-h^2/2M)[d^2/dr^2 - l(l+1)/r^2] u - C(1/r)[1 - (1 - \exp(-\kappa r))]u = Eu \qquad (13)$$

At C<0, the screened potential is split into a Coulomb part (~1/r) and a reverse-sign part ~ $-(1/r)[1 - \exp(-\kappa r)] < 0$. Consequently, the screened potential at C<0 gives rise to a parallel binding potential which is only vanishing at the singularity point at r = 0. Its maximum 1 is reached at $\kappa \gg 1/r_0$ or at $r = \infty$. This parallel binding potential decreases the scattering potential of two Coulomb particles. As the screening concentration n is increased, the binding power increases too to completely compensate for the scattering at very large $\kappa_S = \sqrt{(8\pi n_0 e^2/\varepsilon k_B T)}$. Unless proven otherwise, this compensating power may be regarded as a concentration-dependent correction $\rho_n = 1 - \exp(-\kappa r)$ to be introduced to the scattering potential in (12) as $-C(1/r)(1 - \rho_n)$. This is justified as long as $n_0$ stands for the statistical bulk average though not the actual concentration of charged defects whose distribution in an external field is given as above. For this reason $n_0$ and $\kappa$, for that matter, should only enter as parameters of the theory based on the Coulomb functions. Formally the $\rho$-parameter should enter into the effective charge q of the electrostatic problem giving $q_{eff}^2 = q^2\rho$.

4. Mathematically, the eigenvalue problem of the screening potential at intermediate carrier concentrations can only be dealt with by solving a nonlinear problem, due to the dependence of $\kappa = \kappa(n)$. Physically, there are two regimes where the problem can be regarded as a single-electron one: (a) at concentrations sufficiently low to warrant that $\kappa r_0 \ll 1$, where $r_0$ is a length characteristic of the Coulomb system and (b) at concentrations sufficiently high to warrant that $\kappa r_0 \gg 1$. In (a) the electron gas is weakly non-ideal and non-degenerate, in (b) the electron gas is weakly non-ideal too but is degenerate [2]. As a reasonable definition of a length we choose $r_0 = e^2/\varepsilon k_B T$; then we get $\kappa = \sqrt{(8\pi n r_0)}$ and thereby $\kappa r_0 = \sqrt{(8\pi n r_0)} r_0 = 1$ reduces to $8\pi n r_0^3$ where $n r_0^3$ is the number of bulk charge carriers within the characteristic volume $v_0 = r_0^3$. Accordingly, (a) and (b) correspond to $n r_0^3$ being very small ($\ll 1$) and quite large ($\gg 1$), respectively. The former case has been considered under 1-2 above. The latter case describes an overcompensated situation ($\exp(-\kappa r) \to 0$) at any finite $r > r_0$ (a strongly-compensated case) corresponding to nearly-free carriers. In the former weakly-compensated case (a) the corresponding radial Schrödinger equation reads as in (10).

### 4. Spherical and cylindrical symmetries

Incorporating Laplace's operator $\Delta_{r,\theta,\varphi}$ in equation (2) and integrating over the angles, we obtain the P-B equation appropriate to spherical symmetry

$$(1/r^2)\,(d/dr)\,(r^2 d\Phi/dr) = \sinh\Phi \tag{14}$$

or, equivalently,

$$d^2\Phi/dr^2 + (2/r)\,d\Phi/dr - \sinh\Phi = 0 \tag{15}$$

The respective spherical symmetry Schrödinger equation reads (applying the substitution $\Psi(\mathbf{r}) = [\psi(r)/r]\,Y_l^m(\theta,\varphi)$ as above):

$$(-h^2/2M)[d^2/dr^2 - l(l+1)/r^2]\psi - \Phi\psi = E\psi \tag{16}$$

The electrostatic screening problem is solved through combining equations (15) and (16). We shall attempt to verify solving equation (14) at $\Phi \ll 1$ by means of $\Phi(r) = (\kappa^2/r)\exp(-\kappa r)$ (Debye's solution) and indeed we find $d^2\Phi/dr^2 + (2/r)d\Phi/dr - \Phi = 0$ for $\Phi(r) = (\kappa^2/r)\exp(-\kappa r)$.

For the complete potential $\sinh\Phi = \frac{1}{2}[\exp\Phi - \exp{-\Phi}] \neq \Phi$ and equation (14) holds good. (However, for $\sinh\Phi \sim \Phi$, the solution is as above.) By substitution $u = r\Phi$ we get $u'' + u = 0$ which is solved in $u(r) = A\exp(-\kappa r)$. (From Ref. [6] equation **2.101**).

In circular cylindrical coordinates, the Laplace operator reads:

$$\Delta_{r,\varphi,z}\,\Phi = d^2\Phi/dr^2 + (1/r)\,(d\Phi/dr) \tag{17}$$

and the PB equation turns in:

$$d^2\Phi/dr^2 + (1/r)\,(d\Phi/dr) - \sinh\Phi = 0 \tag{18}$$

Its linearized version is solved in cylindrical Bessel functions: $\Phi(r) \propto J_0(2\sqrt{[\lambda r]})$, whereas the eigenvalues obtain from $J_0(2\sqrt{\lambda}) = 0$. (From Ref. [6] equation **2.95**). The solutions to various symmetry forms of the screening equations (P-B or D-H) are summarized in Table I. Note that there are no periodic D-H solutions, a genuine feature of the complete screening equation.

Table I
Solutions to the electrostatic problem (Ref.[3,7,8])

| Symmetry | DH solution | Periodic PB solution | Aperiodic solution |
|---|---|---|---|
| spherical | $(1/r)\exp(-\kappa r)$ | $2\ln[\cotan(\frac{1}{2}\kappa r)]$ (asymptotic) | $2\ln[\cotanh(\frac{1}{2}\kappa r)]$ |
| cylindrical | $J_0(2\sqrt{[\lambda r]})$ | $2\ln[\cotan(\frac{1}{2}\kappa r)]$ (asymptotic) | $\ln\{[(\kappa r)^2\sin^2(\frac{1}{2}\sqrt{H}\times \ln(r/r_0)]/H\}$ |
| 1-D | $\propto \exp(-\kappa x)$ | $2\ln[\cotan(\frac{1}{2}\kappa x)]$ (exact) | $2\ln[\cotanh(\frac{1}{2}\kappa x)]$ (exact) |

## 5. Asymptotic solution forms

In this connection we are interested in the center-close ($r \ll r_0$) and asymptotic ($r \gg r_0$) forms of the spherical and cylindrical symmetries of the complete PB equation. We set for the asymptotic forms ($r \gg r_0$) ($r_0$ – the core radius of a dislocation):

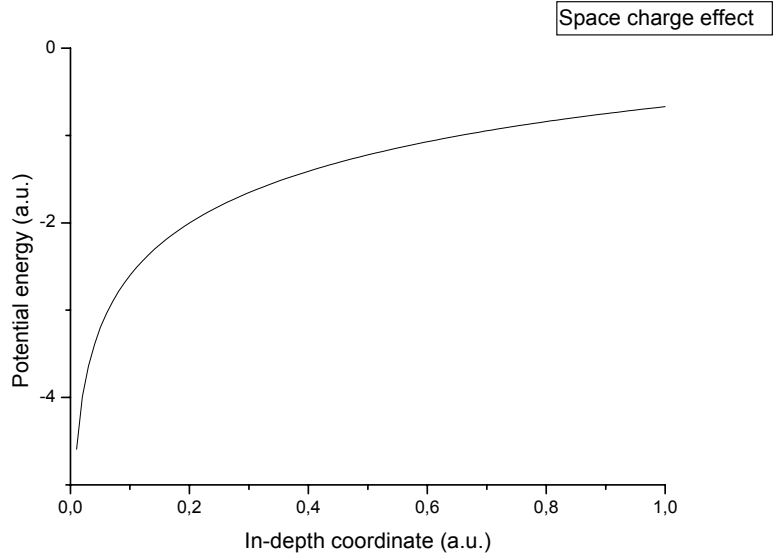

Figure 1: The space charge layer which screens a charged surface from the bulk, according to the simple electrostatic theory [11] using equations of the last row in Table I.

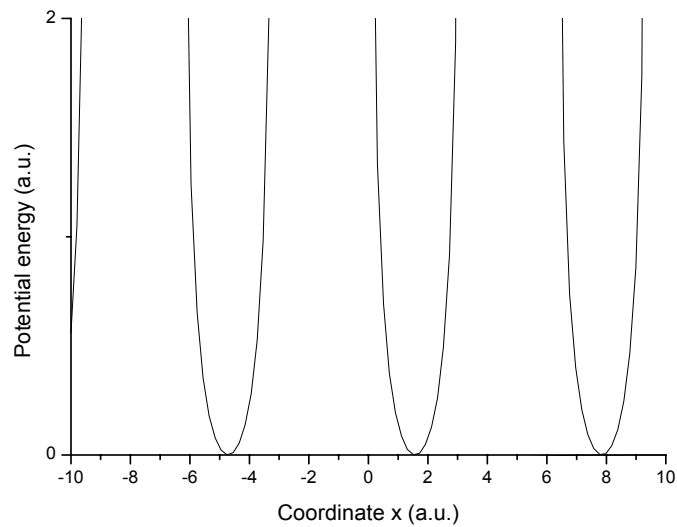

Figure 2: The periodic potential from the nonlinear PB equation. Such potentials describe wave-like space-charge arrangements. Barriers are formed between dagger-like regions [3,7].

$\Phi'' + (2/r)\,\Phi' - \sinh\Phi = 0$ (spherical) $\Phi'' - \sinh\Phi \propto 0$ $\Phi = \ln[\cotan(½\,\kappa r)]^2$ (periodic)

$\Phi = \ln[\cotanh(½\,\kappa r)]^2$ (aperiodic)

(19)

$\Phi'' + (1/r)\,\Phi' - \sinh\Phi = 0$ (cylindrical) $\Phi'' - \sinh\Phi \propto 0$ $\Phi = \ln[\cotan(½\,\kappa r)]^2$ (periodic)

$\Phi = \ln[\cotanh(½\,\kappa r)]^2$ (aperiodic)

Curiously, from an asymptotic distance the r = 0 line and the point source give rise to similar patterns, as exemplified in Figure 3. For the center-close forms we set:

$\Phi'' + (2/r)\,\Phi' - \sinh\Phi = 0$ $(2/r)\,\Phi' - \sinh\Phi \sim 0$ (spherical) $\ln \tanh(½\Phi) = ¼(\kappa r)^2$

$\Phi = 2\tanh^{-1}[\exp(¼(\kappa r)^2)] = \ln\{[1 + \exp(¼(\kappa r)^2)] / [1 - \exp(¼(\kappa r)^2)]\}$

(20)

$\Phi'' + (1/r)\,\Phi' - \sinh\Phi = 0$ $(1/r)\,\Phi' - \sinh\Phi \sim 0$ (cylindrical) $\ln \tanh(½\Phi) = ½(\kappa r)^2$

$\Phi = 2\tanh^{-1}[\exp(½(\kappa r)^2)] = \ln\{[1 + \exp(½(\kappa r)^2)] / [1 - \exp(½(\kappa r)^2)]\}$

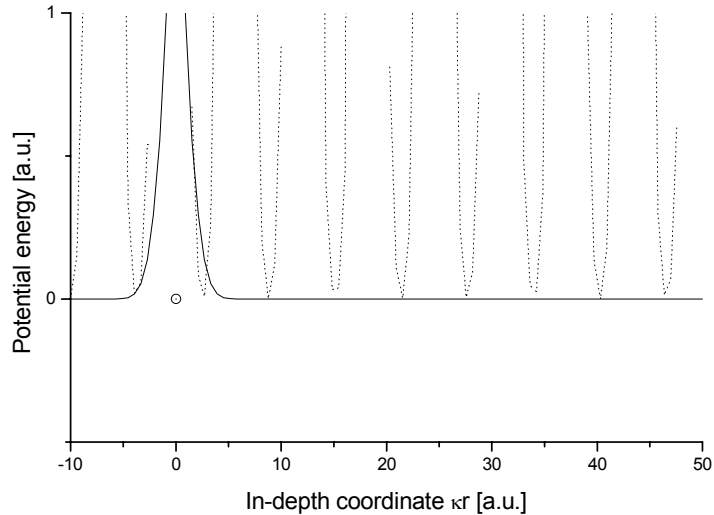

Figure 3: Distribution of periodic potentials (dotted) and the aperiodic subsurface potential (solid) in a crystal as obtained from the nonlinear radial PB equation. In reality, the aperiodic pattern begins just beneath the surface to be replaced at some depth by the periodic feature.

It is remarkable that while the asymptotic behavior may be periodic in κr at large distances r, it is aperiodic at short distances from the source at r = 0 [7]. The resulting forms of solution for the two symmetries are depicted in Figure 3. The cylindrical symmetry PB equation has been solved and applied to the problem of precipitating impurities along dislocation lines [8]. The related bulk electrostatic potential has been found by solving the complete nonlinear problem. No solution is available for the analogous spherical symmetry. However, numerical solutions for two species with spherical symmetry have been reported [9].

The authors of Ref. [8] have split the entire $\Phi$ range into three subranges:

*Small range* $\Phi \ll 1$, where

$$\Delta\Phi \approx \kappa^2\Phi (1 + 1/6\, \Phi^2 + 1/120\, \Phi^4). \tag{21}$$

The solution in this small $\Phi$ range is

$$\Phi(\kappa_{eff}\, r) = -AJ_0(\kappa_{eff}\, r) \tag{22}$$

in which $\kappa_{eff}^2 = \kappa^2(1 + 1/6\, \Phi^2 + 1/120\, \Phi^4)$, $J_0$ is the modified Bessel function of zeroth order, A is a constant.

*Intermediate range* $\Phi \gg 1$, where

$$\Phi'' + (1/r)\Phi' = -\tfrac{1}{2}\exp(-\Phi) \tag{23}$$

whose solution is [8]:

$$\Phi(r) = \ln\{[(\kappa r)^2 \sin^2(\tfrac{1}{2}\sqrt{H}\ln(r/r_0))]/H\} \tag{24}$$

The third region is where the solutions in small and intermediate r are fitted together.

## 6. Schrödinger equation

Once solutions to the P-B equations have been found, they can be inserted into the Schrodinger equation as in eqn. (16) to derive quantum mechanical eigenstates and eigenvalues. Such have been sought and obtained for the minimum-frequency periodic solution (cf. eqn. (20)) [10]. Using Hill's determinant method, several energy bands and energy bandgaps have been deduced at reasonable values of the parameters which bands arise in 1-D periodic structures by intrinsic and/or impurity defects. The occurrence of meaningful energy bands in periodic Maxwell-Boltzmann structures by point defects is due to the area under the periodic potential being permeable by the charge carriers. The magnitude of surface charges and the distribution of compensating space charges in ionic crystals have been obtained and studied using the complete P-B equation for a planar geometry [11]. From the present study it follows that charge carriers in both spherical and cylindrical symmetries can only traverse the corresponding space-charge layers if they have quantized energies falling within one of the point defect bands.

It may be informative to reproduce the obtained eigenvalue relation for the point defect bands ($\kappa = \kappa_S$):

$$\sin^2(\pi k/\kappa) = \Delta(0)\sin^2[\pi\sqrt{(2mE')}/h\kappa] \tag{25}$$

where $E' = (\hbar k)^2/2m$, $A = 4mk_BT/\hbar\kappa$, $\Delta(0) = \det\|A/(n^2 - C_0)\|$ is Hill's determinant [12]. If $\Delta(0) = 1$, then $E = E'$ as for a free particle. However, if $\Delta(0) \neq 1$ equation (26) can be satisfied for real E and k only when E is quantized in allowed energy bands.

We also note that eqn. (4) gives the least-frequency periodic solution of the P-B equation which solution is thermodynamically the most favorable one in terms of the free energy [7]. This underlines the general significance of the conclusion for the existence of allowed energy bands in the traverse of charged defects across space-charge layers beneath the surface of (ionic) crystals.

It should be stressed finally that References [2] and [10] are the only ones so far to deal with the changes of the quantized energy spectrum of crystal induced by charge-screening electrostatic potentials, Poisson-Boltzmann or Debye-Hückel ones. These studies are also informative for the nature of the short-range potentials suggested by Yukawa to represent a phenomenological approach to the inter-nuclear interactions. Now, if Yukawa's model is extended to the level of the P-B nonlinear status, then it would suggest the occurrence of a periodic shell-like structure within a nucleus, a model discussed earlier here and there in relevance to nuclear matter.